\documentclass[sigconf]{acmart}
\AtBeginDocument{%
  }

\usepackage{multirow}
\usepackage{graphicx}
\usepackage[english]{babel}
\usepackage{enumitem}
\usepackage{listings}
\usepackage{xcolor}
\usepackage{amsmath,mathrsfs}
\usepackage{tcolorbox}
\usepackage{enumitem} 
\usepackage{ifthen}
\usepackage{hyperref}
\usepackage{subcaption}
\hypersetup{
    colorlinks=true,
    linkcolor=blue,
    filecolor=magenta,      
    urlcolor=blue,
    }

\usepackage{svg}

\newcommand{\sigir}[1]{\textcolor{black}{#1}}

\addto\extrasenglish{%
}
\copyrightyear{2026}
\acmYear{2026}
\setcopyright{cc}
\setcctype{by-nc-nd}
\acmConference[SIGIR '26]{Proceedings of the 49th International ACM SIGIR Conference on Research and Development in Information Retrieval}{July 20--24, 2026}{Melbourne, VIC, Australia}
\acmBooktitle{Proceedings of the 49th International ACM SIGIR Conference on Research and Development in Information Retrieval (SIGIR '26), July 20--24, 2026, Melbourne, VIC, Australia}
\acmDOI{10.1145/3805712.3808370}
\acmISBN{979-8-4007-2599-9/2026/07}
\begin{document}

\title[TRACE: Tourism
Recommendation with Agentic Counterfactual Explanations]{TRACE: A Conversational Framework for Sustainable Tourism Recommendation with Agentic Counterfactual Explanations}

\author{Ashmi Banerjee}
\email{ashmi.banerjee@tum.de}
\affiliation{%
  \institution{Technical University of Munich}
  \city{Munich}
  \country{Germany}
}
\author{Adithi Satish}
\email{adithi.satish@tum.de}
\affiliation{%
  \institution{Technical University of Munich}
  \city{Munich}
  \country{Germany}
}

\author{Wolfgang W\"orndl}
\email{woerndl@in.tum.de}
\affiliation{%
  \institution{Technical University of Munich}
  \city{Munich}
  \country{Germany}
}

\author{Yashar Deldjoo}
\email{yashar.deldjoo@poliba.it}
\affiliation{%
  \institution{Polytechnic University of Bari}
  \city{Bari}
  \country{Italy}
}

\renewcommand{\shortauthors}{Ashmi Banerjee, Adithi Satish, Wolfgang Wörndl, and Yashar Deldjoo}

\begin{abstract}
Traditional conversational travel recommender systems primarily optimize for user relevance and convenience, often reinforcing popular, overcrowded destinations and carbon-intensive travel choices. To address this, we present \textbf{TRACE} (Tourism Recommendation with Agentic Counterfactual Explanations), a multi-agent, LLM-based framework that promotes sustainable tourism through interactive nudging. TRACE uses a modular orchestrator-worker architecture where specialized agents elicit latent sustainability preferences, construct structured user personas, and generate recommendations that balance relevance with environmental impact. A key innovation lies in its use of \textit{agentic counterfactual explanations} and LLM-driven clarifying questions, which together surface greener alternatives and refine understanding of intent, fostering user reflection without coercion. User studies and semantic alignment analyses demonstrate that TRACE effectively supports sustainable decision-making while preserving recommendation quality and interactive responsiveness.
TRACE is implemented on Google’s Agent Development Kit, with full code, Docker setup, prompts, and a publicly available demo video to ensure reproducibility. 
A project summary, including all resources, prompts, and demo access, is available at \url{https://ashmibanerjee.github.io/trace-chatbot}.

\end{abstract}



\maketitle

\section{Introduction}

Conversational agents are increasingly used to support complex decision-making tasks such as travel planning. Recent advances in LLMs have enabled chat-based travel assistants that can flexibly respond to natural language queries and generate personalized recommendations~\cite{wang2025toward, gu2024survey, shao2025personal}. However, most existing Conversational Recommender Systems (CRS) for travel primarily optimize for user relevance and convenience, often reinforcing popular, overcrowded destinations and carbon-intensive travel choices. As a result, sustainability considerations, such as environmental impact, congestion, and seasonality, are typically treated as secondary constraints, if considered at all~\cite{banerjee2023fairness,fang2024macrs,macrec2024}.

In this demo paper, we present \textbf{TRACE} (\emph{Tourism Recommendation with Agentic Counterfactual Explanations}), an LLM-agent-based conversational recommender system explicitly designed to promote sustainable tourism practices. Unlike traditional chatbots or chat-based travel assistants, TRACE incorporates sustainability objectives from the ground up. The system prioritizes recommendations such as less-visited or emerging destinations over overcrowded hotspots, and favors lower-impact transportation options (e.g., train travel over flights) whenever feasible. Crucially, these sustainability signals are not hard-coded defaults but are inferred through interaction.

The primary goal of this work is \emph{not} to build a state-of-the-art recommender system in terms of predictive accuracy or ranking performance. Instead, we aim to explore the design space and research opportunities enabled by multi-agent LLM architectures for sustainable conversational recommendation. In particular, we focus on how such systems can be structured to \emph{nudge} users toward more sustainable travel choices without violating their explicit preferences or degrading user trust.

TRACE achieves this through a modular, agentic design. Dedicated agents generate targeted clarifying questions to elicit latent sustainability preferences, construct structured user personas, and produce recommendations that jointly optimize for relevance and sustainability signals. A key contribution of the system is its explanation agent, which generates persuasive justifications highlighting the sustainability benefits of recommended options. When users do not explicitly express sustainability preferences, the system employs counterfactual explanations~\cite{guidotti2024counterfactual} to gently expose greener alternatives, enabling informed reflection rather than coercive intervention.

We demonstrate TRACE as an interactive system that showcases how agentic LLM pipelines can operationalize responsible nudging in tourism recommendations. By releasing this demo, we aim to stimulate discussion on the role of conversational interfaces, explanations, and counterfactual reasoning in shaping more sustainable user behavior and to encourage further research beyond accuracy-centric evaluation paradigms in recommender systems~\footnote{Demo video: \url{https://youtu.be/BdtEFSp42fw}}. 

\paragraph{Contributions}
This paper introduces TRACE, a multi-agent, LLM-based conversational recommender system for sustainable tourism:
\begin{itemize}
  \item \textbf{Modular, multi-agent framework}: Specialized agents handle user modeling, clarifying questions, recommendations, and explanations, integrating digital nudging to promote greener choices.
  \item \textbf{Open-source release and demo}: The full code and framework are publicly available, accompanied by a demo video to facilitate reproducibility and further research~\footnote{\url{https://ashmibanerjee.github.io/trace-chatbot}}.
  \item \textbf{Empirical evaluation}: User studies and semantic metrics show TRACE balances relevance, and sustainability nudging while maintaining interactive response times.
\end{itemize}

\paragraph{Related Work}

Conversational recommender systems have been widely studied for interactive preference elicitation and natural language decision support. Recent LLM-powered, multi-agent approaches leverage specialized agents for user modeling, planning, and explanation, enabling collaborative recommendation and flexible reasoning~\cite{huang2025towards, maragheh2025future,llmagentsurvey2025}. However, these systems primarily focus on accuracy and engagement, leaving sustainability and behavior change as secondary concerns~\cite{fang2024macrs,macrec2024}. 
Although recent work~\cite{banerjee2025collab} has explored multi-agent frameworks to balance relevance and sustainability in tourism recommendations, these approaches do not explicitly incorporate conversational interactions or leverage explanations and counterfactuals to nudge user behavior, and they remain primarily research prototypes with limited production readiness.
Explainable and counterfactual recommendation research further explores transparency and bias mitigation, often via minimally contrastive explanations~\cite{cecr2024,lxr2024,tan2021counterfactual,wang2024counterfactual, guidotti2024counterfactual}. Relatedly, clarifying questions have been investigated as a way to uncover hidden or incomplete user preferences~\cite{ren2021learning,ivan2021towards,sekulic2024estimating,bi2021asking}, with taxonomies spanning vague, spatial, and temporal dimensions~\cite{zamani2020generating} and LLM-based dialogue updates supporting personalization~\cite{kemper2024retrieval}. In parallel, sustainability-aware recommenders apply digital nudging and eco-objectives to guide greener choices~\cite{mauro2024point,halimeh2025towards,banerjee2024green,banerjee2025smartsustain, banerjee2025modeling}, though most remain single-agent, non-conversational, rely on static sustainability encodings, and often do not leverage LLMs. In contrast, \textbf{TRACE} integrates these directions in a multi-agent, LLM-based conversational framework, where agents infer preferences, build structured personas, and explain greener choices as an evolving dialogue objective.

\begin{figure}[htbp]
    \centering
      \includegraphics[width=\linewidth,trim={0.5cm 0 1.5cm 0},clip]{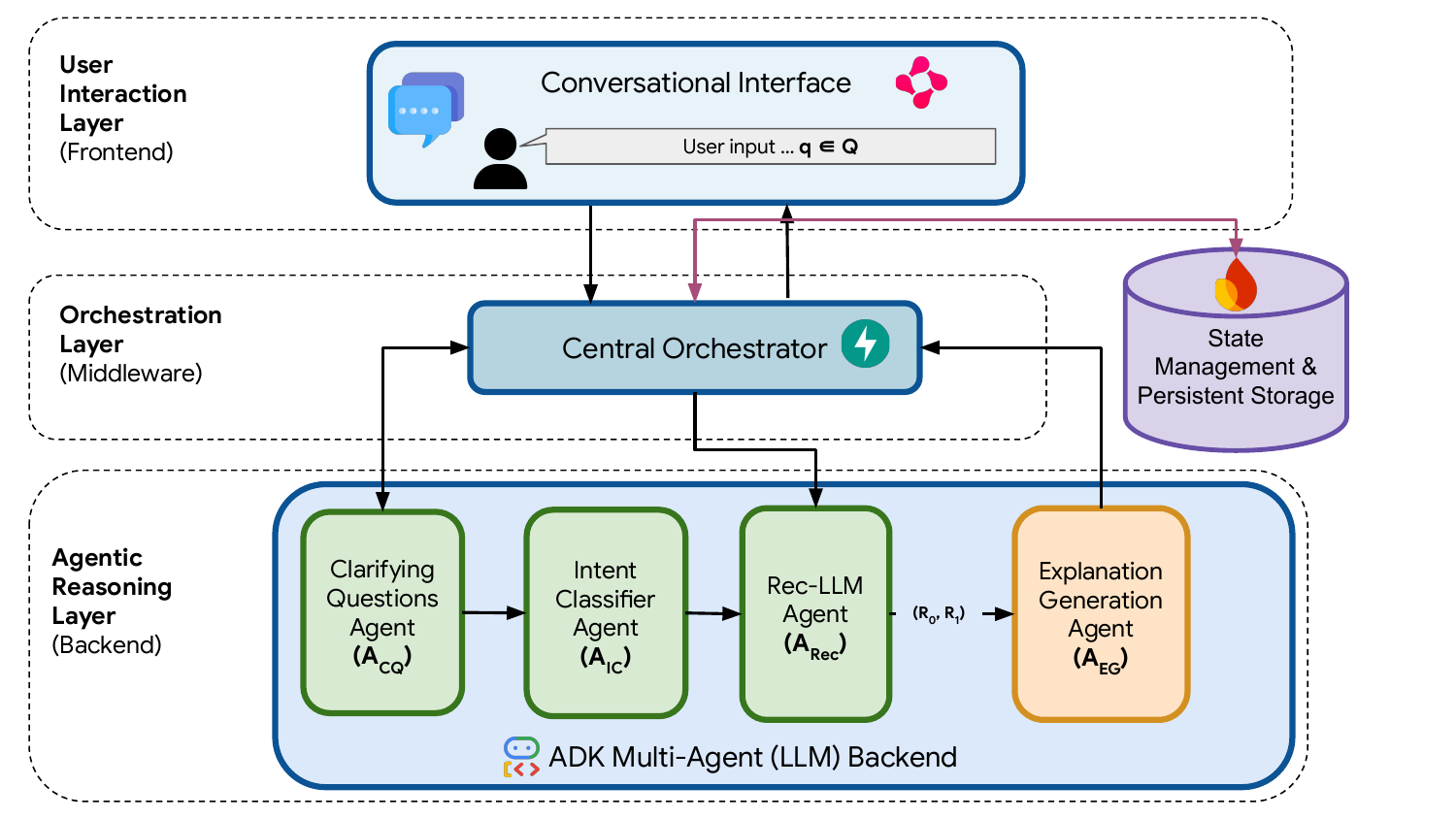}
    \caption{System Architecture of the CRS Chatbot. This diagram illustrates the Orchestrator-Worker paradigm, detailing the flow between the User Interaction layer (Frontend), Orchestration layer (Middleware), and the Agentic Reasoning layer (Backend).}
    \label{fig:system_architecture}
\end{figure}

\section{TRACE Framework: System Design}

The proposed system adopts a modular, microservices-based architecture for delivering sustainable travel recommendations using a Multi-Agent LLM framework. The design follows an Orchestrator-Worker paradigm, where a central middleware component coordinates user interactions and delegates reasoning tasks to specialized agents.
The architecture is organized into three conceptual layers:
\begin{itemize}
    \item \textbf{User Interaction Layer (Frontend)}, responsible for conversational interaction and feedback collection.
    \item \textbf{Orchestration Layer (Middleware)}, which manages session state, control flow, and agent execution.
    \item \textbf{Agentic Reasoning Layer (Backend)}, consisting of a sequential pipeline of specialized LLM agents implemented using Google's Agent Development Kit (ADK) over vertexAI.

\end{itemize}

Throughout this paper, we define \emph{context} as the information accumulated from a user’s initial query and responses to clarifying questions, which collectively inform the recommendation process by capturing inferred preferences and sustainability considerations.
This layered design promotes modularity, scalability, and clear separation of concerns between interaction handling, control logic, and reasoning components.
Figure~\ref{fig:system_architecture} illustrates the overall system architecture, highlighting the interactions between the user interface, orchestrator, and various LLM agents.

\subsection{Formalization of the Multi-Agent CRS}

We formalize the TRACE Conversational Recommender System (CRS) as a sequential pipeline of transformations over the state space of a user session. Let $q \in \mathcal{Q}$ denote the initial natural language query. The system proceeds through the following stages.

\subsubsection{Elicitation and Persona Modeling}

The Clarifying Question Agent $A_{\mathrm{CQ}}$ generates a set of targeted questions
$C = \{c_1, c_2, \dots, c_n\}$ to elicit latent user preferences related to sustainable travel:
\begin{equation}
    C = A_{\mathrm{CQ}}(q).
\end{equation}

The primary goal of the Clarifying Question Agent is to elicit the user's sustainability preferences indirectly and their willingness to compromise, for example, by asking whether they would be willing to travel to a lesser-known place rather than a popular (and crowded) destination, thereby improving intent detection. However, if the user's initial query is vague (\textit{"I want to travel in Europe"}), then $A_{\mathrm{CQ}}$ is also instructed to ask the user about general travel aspects such as their budget and interests.
The user's initial query is provided as input to $A_{\mathrm{CQ}}$, which generates up to 5 clarifying questions, which are then shown to the user one at a time.

Given the user responses $\Gamma$ to these questions, the Intent Classification Agent $A_{\mathrm{IC}}$ maps the interaction into a structured \emph{User Travel Persona} $\mathcal{U}$ and a vector representing the user’s \emph{Willingness to Compromise} (WTC) across sustainability dimensions such as emissions, congestion, and seasonality:
\begin{equation}
    (\mathcal{U}, \mathrm{WTC}) = A_{\mathrm{IC}}(\Gamma, q).
\end{equation}

The generated \textit{User Travel Persona} and \textit{Willingness to Compromise} are provided to the recommender to obtain personalized, context-aware recommendations.

\subsubsection{Recommendation Generation}

The system constructs two candidate recommendation sets. The baseline set $R_0$ is instructed to prioritize relevance solely to the original query $q$. In contrast, the sustainable set $R_1$ incorporates the inferred user persona and sustainability preferences from $A_{\mathrm{IC}}$. 
Both sets are generated by the Recommender Agent $A_{\mathrm{Rec}}$, a Rec-LLM using few-shot prompting. In particular, $R_1$ is produced by prioritizing sustainability-related signals $S$ derived from the clarification responses $\Gamma$, while preserving alignment with the inferred persona $\mathcal{U}$:
\begin{equation}
    R_1 = A_{\mathrm{Rec}}(\mathcal{U}, \Gamma, S).
    \label{eq:rec}
\end{equation}

\subsubsection{Persuasive Explanation and Decision Logic}

The Explanation Generation Agent $A_{\mathrm{EG}}$ serves as the final decision-making component. It outputs a selected recommendation $r^*$, a persuasive explanation $E$, and a counterfactual alternative $r_{\mathrm{alt}}$, explicitly conditioned on the inferred willingness to compromise:
\begin{equation}
    (r^*, E, r_{\mathrm{alt}}) =
    A_{\mathrm{EG}}(R_0, R_1, \mathcal{U}, \mathrm{WTC}).
    \label{eq:explanation}
\end{equation}

The agent dynamically adopts one of two rhetorical strategies:

\paragraph{Direct Alignment}
If $\mathrm{WTC}$ indicates openness to sustainability trade-offs, the agent selects $r^* \in R_1$ and generates an explanation emphasizing the sustainability improvement
\[
\Delta S = M(R_1) - M(R_0),
\]
where $M(\cdot)$ denotes a vector of sustainability metrics (e.g., CO$_2$ emissions, visitor pressure, walkability).

\paragraph{Counterfactual Nudging}
If $\mathrm{WTC}$ indicates resistance to sustainability trade-offs, the agent selects the baseline recommendation $r^* \in R_0$ to preserve user trust, while generating a counterfactual explanation for an alternative $r_{\mathrm{alt}} \in R_1$. The explanation is framed conditionally, for example:
\emph{``Had you expressed interest in lower environmental impact, $r_{\mathrm{alt}}$ would have been recommended because \dots''}
This strategy enables implicit nudging toward sustainable options without violating the user's explicitly stated constraints.
\subsection{Implementation}
TRACE is implemented as a modular, state-aware pipeline using \textbf{Google's Agent Development Kit (ADK)}~\cite{google_adk} on Vertex AI~\cite{google_vertex_ai}, with the \texttt{gemini-2.5-flash} model handling all agentic reasoning~\cite{google_gemini_v2_5_2024}. The backend uses an Orchestrator-Worker pattern managed via \textbf{FastAPI}~\cite{fastapi}, with \textbf{Google Firestore}~\cite{firebase_firestore} maintaining persistent session state (user queries, persona vectors, and $WTC$).  
The user interface, built with \textbf{Chainlit}~\cite{chainlit}, supports a seamless Clarifying Question loop and the rendering of recommendations. The stack is containerized with \textbf{Docker} and deployed on \textbf{Google Cloud Run}~\cite{google_cloudrun} for scalable, serverless execution of asynchronous multi-agent workflows. 

\begin{figure}[htbp]
    \centering
    \includegraphics[width=\linewidth]{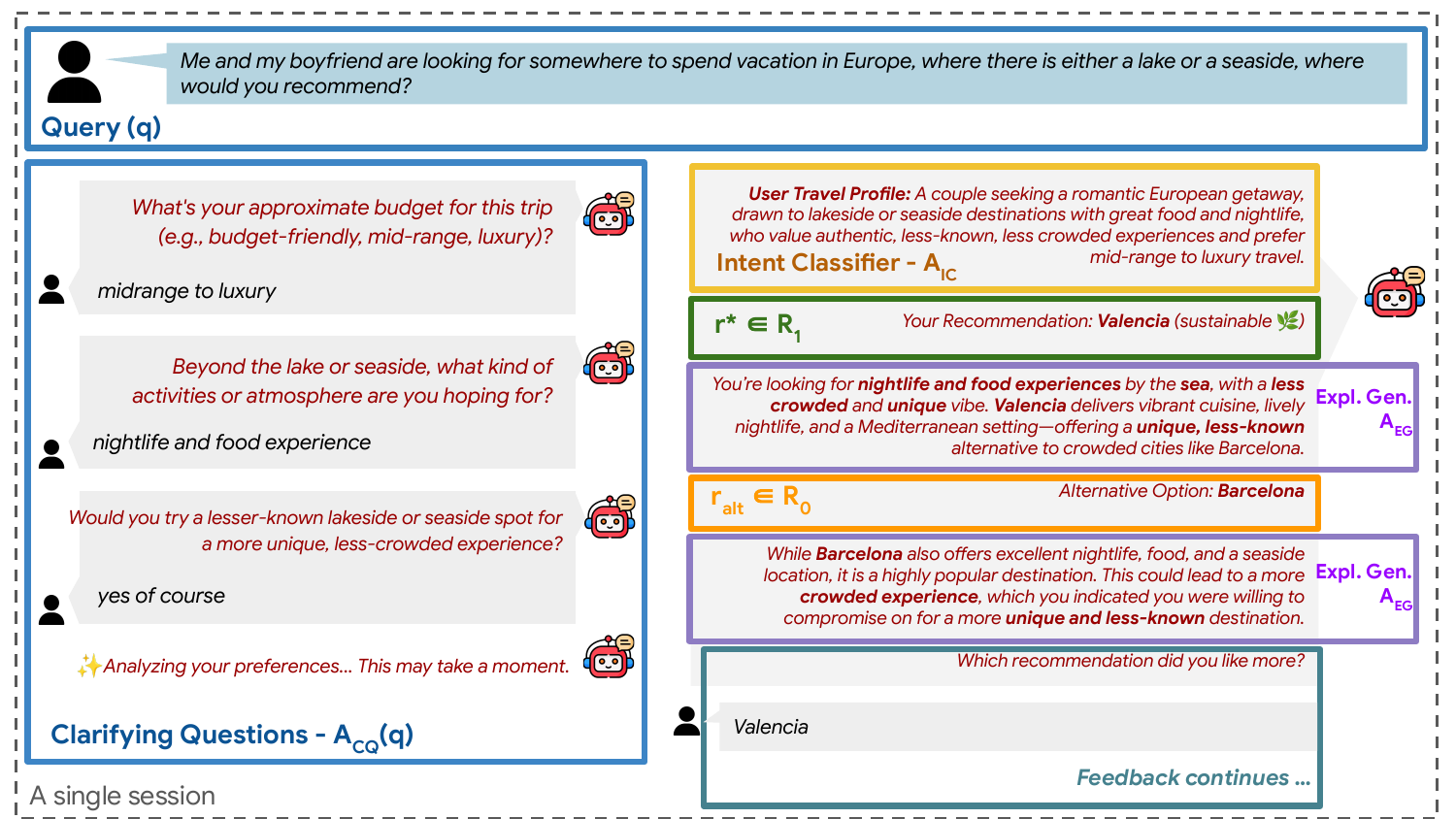}
    \caption{An example TRACE session. The session starts with the user’s initial query, followed by clarifying questions. It concludes with a generated travel profile and a set of recommendations, each accompanied by explanations. The session also collects session-specific feedback, including which recommendations were preferred, how helpful the explanations were, how effective the clarifying questions were, and an option to provide qualitative feedback.}
    \label{fig:example_session}
\end{figure}

Users can enter free-text queries or select predefined travel scenarios that cover a range of preferences and sustainability considerations (e.g., eco-friendly stays, off-the-beaten-path trips). We use sample queries from SynthTRIPS~\cite{banerjee2025synthtrips}, which provide synthetic travel queries and preferences, to offer users inspiration for their initial inputs.  

To prevent misuse, the Clarifying Question Agent ($A_\mathrm{CQ}$) enforces guardrails with few-shot instructions, restricting recommendations for single European city trips. 
Queries outside this scope (for example, \textit{"Recommend some movies to watch this weekend"}) are flagged as invalid, and users are prompted to stay within city trip recommendations. 

\subsection{User Journey}

In this paper, a \emph{session} denotes a complete interaction between a user and the system, beginning with the user’s initial query and ending with their responses to the feedback questions.

Each session proceeds as follows. Given their query, the user first responds to a set of clarifying questions to construct a travel persona. The system then generates two candidate recommendations: a baseline recommendation $R_0$ and a context-aware recommendation $R_1$. Based on the user’s willingness to compromise (WTC) on sustainability, the Explanation Generation Agent determines which candidate is presented as the primary recommendation, $r^*$, and which is shown as the alternative, $r_{\mathrm{alt}}$. When the user indicates no willingness to compromise on sustainability, the agent may assign the baseline recommendation ($R_0$) as the primary option.

\autoref{fig:example_session} illustrates an example session in which the user expresses a preference for a seaside holiday. In this instance, openness to exploring lesser-known destinations to avoid crowds results in \textit{Valencia} ($R_1$) being presented as the primary recommendation ($r^*$), with \textit{Barcelona} ($R_0$) shown as the alternative ($r_{\mathrm{alt}}$). Had the user instead preferred popular destinations without such flexibility, the roles of the two recommendations would have been reversed.

\section{Evaluation and User Insights}
We evaluate TRACE using a user study and quantitative alignment analysis to assess its ability to promote sustainable travel choices and support coherent agentic reasoning.

\subsection{User Study Design}
We recruited 24 participants (58.3\% male, 41.7\% female) via social media, personal networks, and university mailing lists. Most participants were aged 18--34 (79\%), and 54\% reported frequent chatbot use. Participants interacted with TRACE using 1--2 travel queries and were encouraged to include challenging or out-of-scope requests (e.g., non-European destinations) to test robustness.

After the interaction, participants rated the system on a 5-point Likert scale (1: Not at all, 5: Extremely)~\cite{joshi2015likert} across three dimensions: \textit{quality of clarifying questions}, \textit{persuasiveness of explanations}, and \textit{extent of choice reconsideration}. After filtering incomplete and out-of-scope responses, we analyzed $N=107$ valid conversations.

The evaluation examines whether TRACE can nudge users toward sustainable choices without being coercive, while maintaining alignment across agents and with user intent. We formalize these objectives through the following research questions (RQs):

\begin{enumerate}
    \item \emph{\textbf{User Feedback and Interaction Analysis (RQ1):} How effectively does the system promote sustainable choices while preserving user trust and engagement?}
    \item \emph{\textbf{System Alignment and Semantic Metrics (RQ2):} To what extent does the system maintain alignment across its internal agents and with user intent?}
    \item \emph{\textbf{System Latency (RQ3):} Does TRACE achieve response times suitable for real-time interaction despite the computational overhead of multiple agents?}

\end{enumerate}

\subsection{RQ1: User Feedback and Interaction Analysis}

\begin{figure}[htbp]
    \centering
    \includegraphics[width=0.9\linewidth]{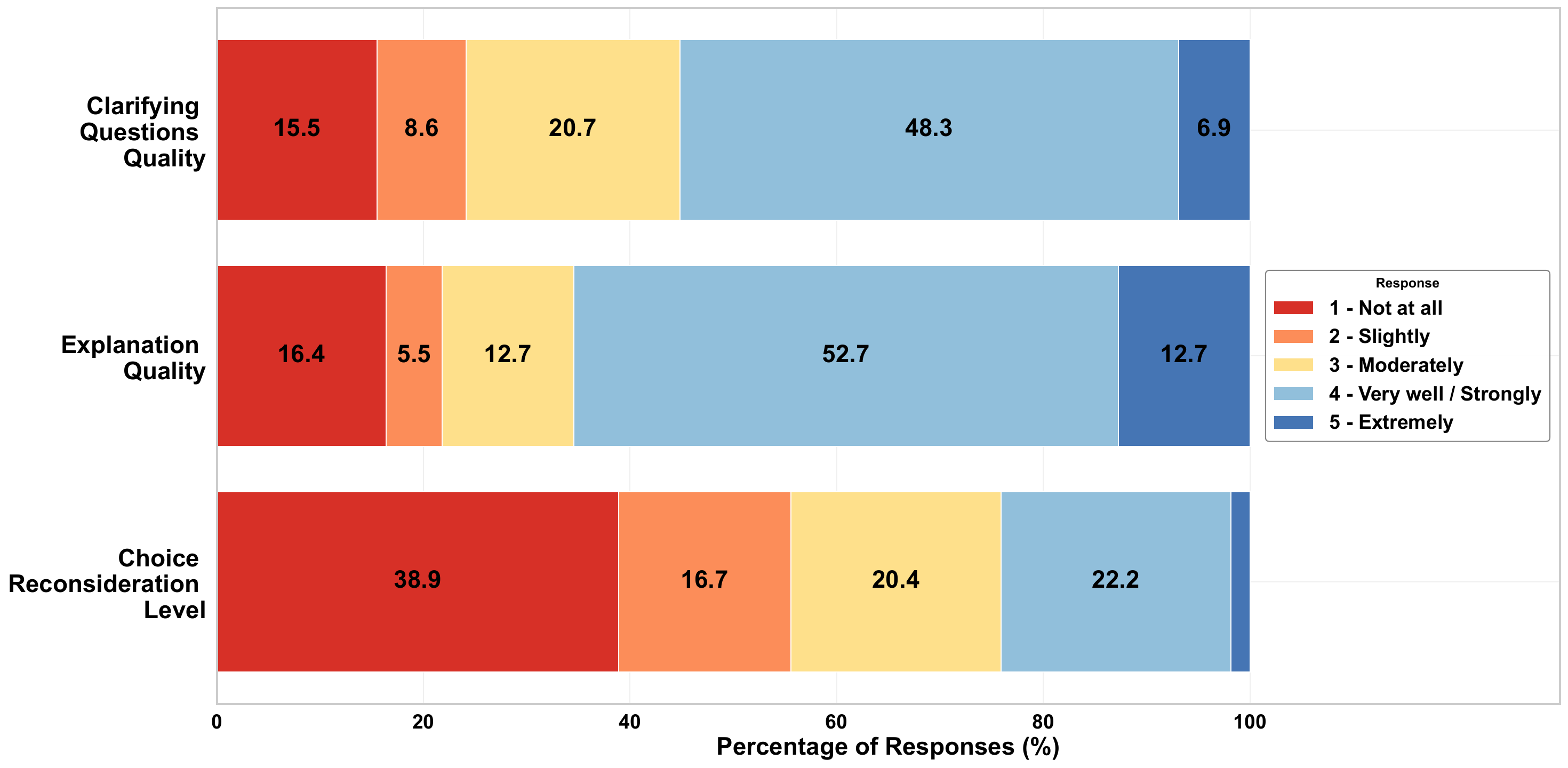}
    \caption{Combined feedback distribution across rating categories for Clarifying Question Quality, Explanation Quality, and Choice Reconsideration Level.}
    \label{fig:feedback}
\end{figure}

RQ1 investigates whether TRACE promotes sustainable choices while preserving user trust and engagement. Results from the user study (Figure~\ref{fig:feedback}) show high acceptance of the system’s conversational interaction. When presented with a primary and alternative recommendation, 79.1\% of users selected the primary option ($r^*$), which in 75.5\% of sessions corresponded to the context-aware (typically more sustainable, i.e., $R_1$) recommendation. In sessions where the baseline ($R_0$) was presented as the primary option ($r^*$), 16.7\% of users selected the context-aware alternative ($R_1$), indicating a modest nudging effect.

User feedback further supports the effectiveness of the interaction design: 55.2\% of users rated the clarifying questions and 65.4\% rated the explanations as \textit{Very Well} or \textit{Extremely Well}. Moreover, around 60\% of users reported some degree of reconsideration of their initial choice, suggesting that TRACE encourages reflection while preserving user agency. Overall, these findings indicate that TRACE can support sustainable recommendations without compromising user engagement or trust.

\subsection{RQ2: System Alignment and Semantic Metrics}

RQ2 examines whether TRACE maintains alignment between the conversation context, the structured user model, and the explanations generated by the agentic pipeline. Using semantic similarity metrics computed with the \texttt{all-MiniLM-L6-v2} model, we measure coherence across these components. The explanations generated by $A_{EG}$ show high similarity to the full conversation (User Query $q$ + Clarifying Questions), with a mean score of 0.7033, indicating strong contextual grounding. The output of the Intent Classifier (User Persona, Travel Intent, and WTC) also aligns closely with the conversation (0.7883), confirming accurate intent capture. Finally, the high similarity between the Intent Classifier output and the generated explanations (0.7437) demonstrates that $A_{EG}$ consistently conditions its reasoning on the inferred user model and sustainability dimensions. Overall, these results indicate strong alignment across TRACE’s internal agents and with user intent.

\subsection{RQ3: System Latency}

We assess whether TRACE can operate within latency bounds suitable for interactive use despite its multi-agent design. Following the clarifying question phase, TRACE requires an average of 23 seconds to generate user profiles, recommendations, and explanations, with a maximum latency of 38 seconds. 
\sigir{These results show that TRACE’s modular architecture does not introduce prohibitive overhead, supporting the practical feasibility of multi-agent conversational recommendation. Moreover, its latency is competitive with that of contemporary multi-agent frameworks~\cite{drammeh2025multi}, which often exceed 40 seconds. Overall, TRACE effectively balances the computational cost of agentic reasoning with the need for accurate and personalized tourism recommendations.}

\section{Conclusion}

We presented \textbf{TRACE}, a multi-agent LLM-based conversational recommender system for sustainable tourism, combining user modeling, clarifying questions, recommendations, and explanations to nudge users toward greener choices while preserving user trust. User studies and semantic metrics show TRACE balances relevance, trust, and sustainability, nudging with interactive response times.  

Due to a Chainlit vulnerability discovered on 21 January, 2026~\cite{securityweek_chainlit_vuln_2026}, the live app is currently offline, but the code, Docker setup, prompts, and a demo video are publicly available for reproducibility. The system ran successfully from January 16–31, 2026, and can be set up locally using our resources. 

Currently limited to single European city travel, TRACE will be extended in future work to multi-city, multi-day itineraries. Overall, it demonstrates the feasibility of combining multi-agent reasoning, LLMs, and sustainability-aware conversational recommendation in a practical, reproducible framework.
\sigir{At the same time, TRACE raises an important \textit{sustainability paradox}. By consistently nudging users toward less popular “hidden gems,” the system may inadvertently create new hotspots, shifting rather than reducing overtourism. Addressing this requires dynamic, adaptive recommendation strategies that account not only for individual preferences but also for destination-level impacts. Future work should therefore incorporate real-time signals, such as destination capacity, environmental indicators, and user feedback, to continuously refine recommendations and mitigate such unintended consequences.}

\section*{GenAI Usage Disclosure}
We used ChatGPT, Claude, and Gemini for code suggestions, and Grammarly for language refinement; all outputs were critically reviewed, and we take full responsibility for the final version.
\section*{Acknowledgments}
We thank the Google AI/ML Developer Programs team for supporting us with Google Cloud Credits.
\bibliographystyle{ACM-Reference-Format}
\bibliography{main}

\end{document}